\documentclass[journal=jacsat,manuscript=article]{achemso}
\usepackage[version=3]{mhchem}
\usepackage{caption}
\usepackage{subcaption}
\usepackage{graphicx}
\usepackage{xcolor}
\usepackage[normalem]{ulem}

\author{Abbas Shiri}
\affiliation{CREOL, The College of Optics \& Photonics, University of Central Florida, Orlando, FL 32816, USA}
\author{Ayman F. Abouraddy}
\affiliation{CREOL, The College of Optics \& Photonics, University of Central Florida, Orlando, FL 32816, USA}
\email{*raddy@creol.ucf.edu}

\title{Severing the link between modal order and group index using hybrid guided space-time modes}

\abbreviations{IR,NMR,UV}

\begin{document}

\begin{tocentry}

Some journals require a graphical entry for the Table of Contents.
This should be laid out ``print ready'' so that the sizing of the
text is correct.

\end{tocentry}

\begin{abstract}
The structure of an optical waveguide determines the characteristics of its guided modes, such as their spatial profile and group index. General features are shared by modes regardless of the waveguiding structure; for example, modal dispersion is inevitable in multimode waveguides, every mode experiences group-velocity dispersion, and higher-order modes are usually slower than their lower-order counterparts. We show here that such trends can be fundamentally altered -- altogether severing the link between modal order and group index hybrid and eliminating dispersion -- by exploiting hybrid guided space-time modes in a planar multimode waveguide. Such modes are confined in one-dimension by the waveguide and in the other by the spatio-temporal spectral structure of the field itself. Direct measurements of the modal group delays confirm that the group index for low-loss, dispersion-free, hybrid space-time modes can be each tuned away from the group index of the conventional mode of same order, and that the transverse size of these hybrid modes can be varied independently of the modal order and group index. These findings are verified in a few-mode planar waveguide consisting of a 25.5-mm-long, 4-$\mu$m-thick silica film deposited on a MgF$_2$ substrate.
\end{abstract}


\section{Introduction}

Guided modes in optical waveguides and fibers -- regardless of the guiding structure -- share inescapable features regarding their \textit{intra}-modal and \textit{inter}-modal dispersion. From a fundamental perspective, these features stem from the boundary conditions imposed on the guided optical field by the confining structure, whether index-guiding \cite{SalehBook07}, a photonic bandgap \cite{Russell03Science,Joannopoulos08Book}, or a plethora of other mechanisms \cite{Duguay86APL,Siegman03JOSAA,Almeida04OL,Siegman07JOSAB,Wang11OL,Pryamikov11OE}. First among these features is that distinct modes at the same wavelength usually have different modal indices, leading to modal dispersion that is particularly exacerbating in light of the burgeoning interest in multimode fibers and waveguides \cite{Li14AOP,Willner15AOP}. Second, higher-order modes typically have higher modal group indices than their lower-order counterparts; i.e., higher-order modes travel slower. Third, guided modes inevitably incur group velocity dispersion (GVD) \cite{Agrawal10Book}. Each of these features can be separately modified: indices of different modes may potentially be equalized by engineering modal degeneracy \cite{Steel01OL}, and GVD can be engineered \cite{Turner06OE} or minimized -- but not altogether eliminated -- via dispersion flattening \cite{Ferrando00OL} or balancing waveguide and material GVD \cite{Agrawal10Book}. These methodologies necessitate modifying the waveguide structure to achieve their goal, which may not always be desirable. Moreover, there is no comprehensive methodology to tackle all of these issues simultaneously.

We recently addressed part of this challenge by introducing a new class of guided optical modes that we have called `hybrid guided space-time (ST) modes' \cite{Shiri20NC}. These are \textit{pulsed} field configurations in single-mode planar waveguides that are nevertheless confined in both transverse dimensions by two distinct mechanisms. The planar waveguide structure confines the field along one dimension; and confinement is enforced along the unconstrained dimension (where light would normally diffract upon free propagation) by introducing a particular spatio-temporal spectral structure into the pulsed field. By combining these two different mechanisms, hybrid guided ST modes (or hybrid ST modes for brevity) exhibit transverse field localization and propagation invariance along the waveguide. The propagation invariance of hybrid ST modes  is the same featured by their freely propagating counterparts; namely ST wave packets \cite{Kondakci16OE,Parker16OE,Kondakci17NP,Wong17ACSP2,Porras17OL,Efremidis17OL,Yessenov19OPN}. By associating each spatial frequency with a particular wavelength according to a prescribed angular-dispersion profile \cite{Donnelly93ProcRSLA}, unique characteristics emerge for ST wave packets, including propagation invariance \cite{Turunen10PO,FigueroaBook14,Kondakci18PRL,Yessenov19OE,Yessenov19PRA}, tunable group velocity in free space \cite{Salo01JOA,Efremidis17OL,Wong17ACSP2,Kondakci19NC} and in transparent dielectrics \cite{Bhaduri19Optica}, anomalous refraction \cite{Bhaduri20NP}, controllable group-velocity dispersion \cite{Sonajalg96OL,Sonajalg97OL,Zamboni03OC,Yessenov21ACSP,Hall21OL3NormalGVD}, self-healing \cite{Kondakci18OL}, space-time Talbot self-imaging \cite{Hall21APLP}, among other useful properties \cite{Yessenov19Optica,Yessenov19OL,Wong20AS,Guo21Light,Wong21OE}. Despite confinement along one dimension by the planar waveguide, hybrid ST modes nevertheless retain these characteristics of their freely propagating counterparts \cite{Shiri20NC}. However, in contrast to conventional guided modes that are identified by two integers (or by one in the case of planar-waveguide modes), hybrid ST modes are identified by an integer index associated with the underlying conventional planar-waveguide mode, and a \textit{continuous} index that characterizes the angular dispersion introduced along the unbounded dimension. Moreover, because these modes propagate invariantly in unpatterned thin films, they have been found to exhibit ultra-low modal losses \cite{Shiri20NC}.

This new degree of freedom allows a hybrid ST mode to sidestep the restrictions imposed on the conventional fundamental mode in the same single-mode waveguide. For example, the group index of the hybrid ST mode can be tuned independently of the waveguide structure by modifying the field structure, whereas the modal index of the fundamental mode at any given wavelength is fixed. Moreover, hybrid guided ST modes are uniquely GVD-free, while the fundmanental mode \textit{must} experience GVD. Indeed, the axial wave number for the hybrid ST mode is \textit{linear} in frequency, so that GVD and \textit{all} higher-order dispersion terms are eliminated over an extended bandwidth. Although hybrid ST modes have so far been investigated only in the context of single-mode planar waveguides, their salutary features would be even more beneficial in the multimode regime. This raises the prospect of equalizing the modal group indices and modal sizes of two (or more) GVD-free hybrid ST modes, which can be critical for maximizing their coupling during nonlinear interactions.

Here we realize hybrid ST modes in a few-mode waveguide in the form of a 4-$\mu$m-thick silica film on a MgF$_2$ substrate. We excite two hybrid ST modes at the same wavelength associated with the first- and second-order planar-waveguide modes, and vary their spatio-temporal spectra to tune their modal group indices $\widetilde{n}_{\mathrm{WG}}$ \textit{independently} of each other. The achieved change in the modal group indices is $\Delta\widetilde{n}_{\mathrm{WG}}\!\sim\!\pm0.6$ ($\sim\pm40\%$ of their initial values). We thus sever the link between the modal order and the associated group index, thereby allowing for group-velocity-matching of different-order modes, or indeed reversing the typical trend by rendering higher-order modes faster than their lower-order counterparts. Not only can the modal group indices for distinct hybrid ST modes be equalized, but their new common group index can be set above or below that of conventional modes, thereby realizing subluminal or superluminal group delays. We confirm these findings by directly measuring the group delay of the hybrid ST modes with respect to that incurred by the conventional pulsed modes. Furthermore, we simultaneously control the modal widths of the hybrid ST modes along the unconstrained dimension at fixed pulse width. Because the spatial and temporal degrees of freedom are correlated in ST fields in general, changing the transverse spatial width necessitates changing the pulse width. However, by exploiting `spectral recycling' as demonstrated in \cite{Hall21PRAspectralRecycling} for freely propagating ST wave packets, we can decouple the spatial and temporal widths but in the context of hybrid ST modes, rendering these two quantities almost independent of each other. 

\section{Theory of hybrid space-time modes}

\subsection{Modal index of hybrid ST modes}

\begin{figure*}[t!]
\centering
\includegraphics[width=17.5cm]{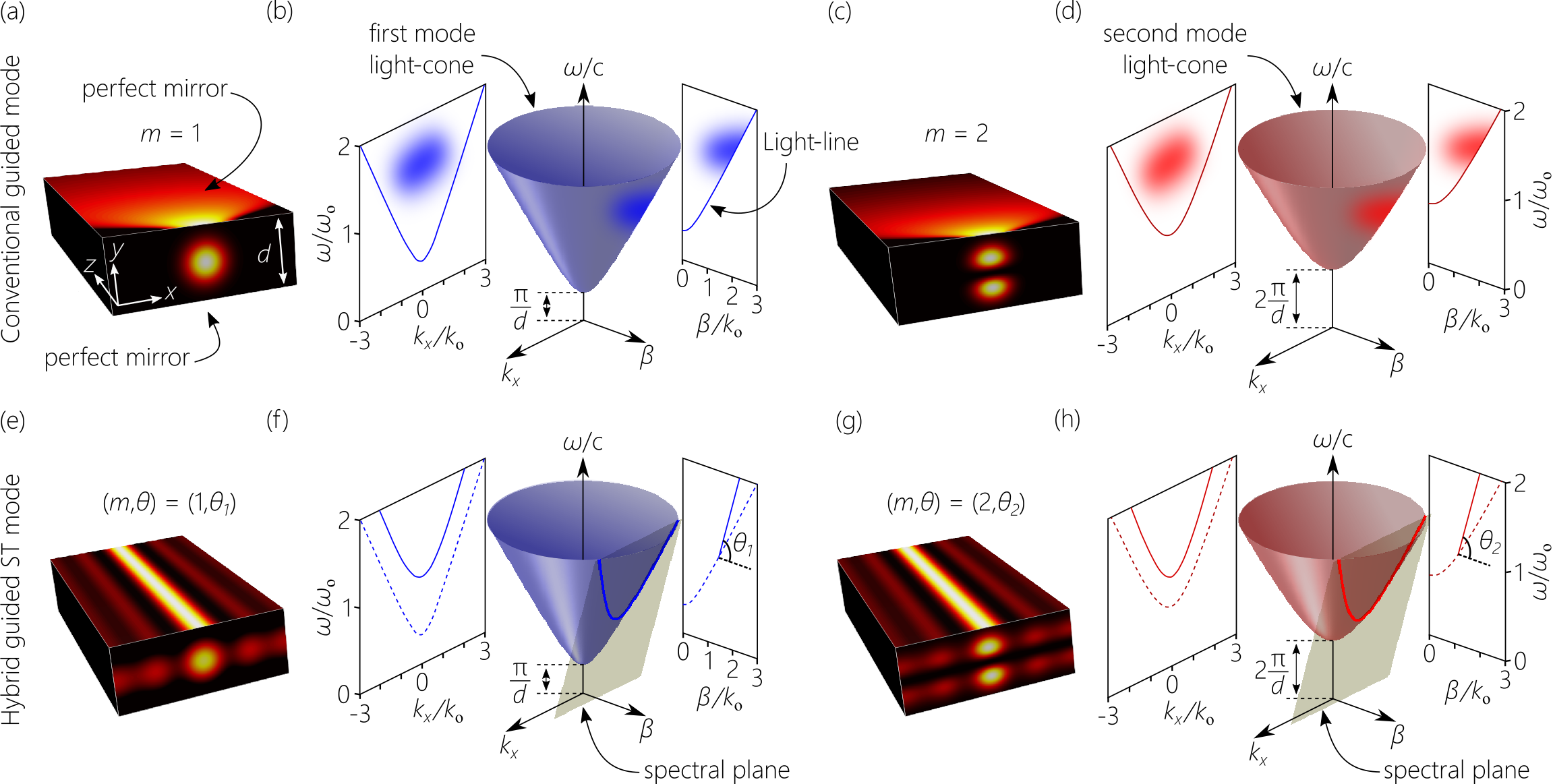}
\caption{\textbf{Comparison between conventional modes and hybrid ST modes in a multimode planar waveguide.} (a) In a planar waveguide formed of a dielectric layer between two perfect mirrors, a focused input field couples to the first-order waveguide mode $m\!=\!1$, remains confined along $y$, but diffracts along the unconstrained dimension $x$. (b) The dispersion relationship for the first-order mode $\beta^{2}+k_{x}^{2}\!=\!(n\tfrac{\omega}{c})^{2}-(\tfrac{\pi}{d})^{2}$. The spatio-temporal spectrum of the focused pulsed input in (a) is a 2D domain on the surface of the modal light-cone. (c,d) Same as (a,b) for the second-order $m\!=\!2$ guided mode. (e) The first-order hybrid ST mode remains invariant while propagating along the planar waveguide, and (f) its spatio-temporal spectrum is restricted to the intersection of the modal light-cone with a spectral plane tilted by an angle $\theta_{1}$ with the $\beta$-axis. The spectral projection onto the $(\beta,\tfrac{\omega}{c})$-plane is a straight line signifying absence of dispersion. (g,h) Same as (e,f) for $m\!=\!2$.}
\label{Fig1}
\end{figure*}

Consider the idealized planar waveguide shown in Fig.~\ref{Fig1}(a) comprising a thin non-dispersive dielectric layer of refractive index $n$ and thickness $d$ between two perfect mirrors. The field is confined along the transverse direction $y$, but is unconstrained along $x$. If the $m^{\mathrm{th}}$-order mode (whether TE or TM) is excited along $y$, then the transverse wave number is $k_{y,m}\!=\!m\tfrac{\pi}{d}$, $m\!=\!1,2,\cdots$, leading to the dispersion relation $\beta^{2}+k_{x}^{2}\!=\!(n\tfrac{\omega}{c})^{2}-(m\tfrac{\pi}{d})^{2}$, where $\beta$ is the axial wave number, $k_{x}$ is the transverse wave number along $x$, $\omega$ is the temporal frequency, and $c$ is the speed of light in vacuum. Geometrically, this dispersion relationship is represented by the surface of a modified cone in $(k_{x},\beta,\tfrac{\omega}{c})$-space that we refer to as the modal light-cone, as depicted in Fig.~\ref{Fig1}(b) for the fundamental mode $m\!=\!1$. The light-line projected onto the $(\beta,\tfrac{\omega}{c})$-plane for $k_{x}\!=\!0$, $\beta^{2}\!=\!(n\tfrac{\omega}{c})^{2}-(\tfrac{\pi}{d})^{2}$, is the dispersion relationship for the fundamental mode of the planar waveguide, which has uniform extent along $x$. Because of the inevitable curvature of the light-line, a pulsed field (of finite temporal bandwidth $\Delta\omega$) will undergo dispersive spreading. Any pulsed field with finite \textit{spatial} extent along $x$ (and, therefore, having finite spatial \textit{and} temporal bandwidths) is represented by a two-dimensional (2D) domain on the surface of the light-cone. The spectral projection onto the $(\beta,\tfrac{\omega}{c})$-plane is also a 2D domain, and thus does not correspond to a propagation-invariant mode. Such a field undergoes diffractive spreading along $x$ in addition to temporal dispersive spreading. Similar observations hold for the $m\!=\!2$ mode, except that the different modal light-cone geometry leads to a different dispersion relationship along the light-line $\beta^{2}\!=\!(n\tfrac{\omega}{c})^{2}-(2\tfrac{\pi}{d})^{2}$; see Fig.~\ref{Fig1}(c,d).

In contrast to these conventional modes, hybrid ST modes are pulsed fields of finite spatial extent [Fig.~\ref{Fig1}(e)] that overcome both diffractive and dispersive spreading in space and time, respectively, by restricting the spectral support domain on the modal light-cone surface to its intersection with a spectral plane defined by the equation $\beta\!=\!nk_{\mathrm{o}}+\tfrac{\Omega}{c}\widetilde{n}_{\mathrm{WG},1}$, where $\Omega\!=\!\omega-\omega_{\mathrm{o}}$, $\omega_{\mathrm{o}}$ is the carrier frequency, $k_{\mathrm{o}}\!=\!\tfrac{\omega_{\mathrm{o}}}{c}$ is the associated wave number, $n_{\mathrm{o}}\!=\!n(\omega_{\mathrm{o}})$, and $\widetilde{n}_{\mathrm{WG},1}$ is the group index of the hybrid ST mode \cite{Shiri20NC} [Fig.~\ref{Fig1}(f)]. This spectral plane is parallel to the $k_{x}$-axis and makes an angle $\theta_{\mathrm{WG},1}$ (the spectral tilt angle) with the $\beta$-axis. The projection of this spectral support domain onto the $(\beta,\tfrac{\omega}{c})$-plane is a straight line making the same angle $\theta_{\mathrm{WG},1}$ with the $\beta$-axis. We show below that it is straightforward to synthesize such a restricted spatio-temporal spectral support domain and to vary $\theta_{\mathrm{WG},1}$. The second-order hybrid ST modes [Fig.~\ref{Fig1}(g)] is constructed in the same manner using the modal light-cone associated with the planar-waveguide mode $m\!=\!2$, with a spectral tilt angle $\theta_{\mathrm{WG},2}$ that can be adjusted independently of $\theta_{\mathrm{WG},1}$. The modal group index of this hybrid ST mode is $\widetilde{n}_{\mathrm{WG},2}\!=\!\cot{\theta_{\mathrm{WG},2}}$. It is important to note that $\widetilde{n}_{\mathrm{WG},1}$ can be tuned independently of the group index $\widetilde{n}_{1}$ of the planar waveguide mode, similarly for $\widetilde{n}_{\mathrm{WG},2}$ and $\widetilde{n}_{2}$.

Because each frequency $\omega$ is associated with a single spatial frequency $\pm k_{x}$ in the spectral projection onto the $(k_{x},\tfrac{\omega}{c})$-plane for a hybrid ST mode, the pulsed field is now endowed with angular dispersion; i.e., each frequency propagates at a different angle $\varphi(\omega)$ with the $z$-axis, such that $k_{x}(\omega)\!=\!n\tfrac{\omega}{c}\sin{\{\varphi(\omega)\}}$. We will see that the unique characteristics of hybrid ST modes stem from the unexpected fact that the derivative for $\varphi(\omega)$ here is not defined at $\omega\!=\!\omega_{\mathrm{o}}$, a condition we refer to as non-differentiable angular dispersion \cite{Hall21OL,Hall21OL3NormalGVD,Hall21OE1NonDiff}.

The spectral projections onto the $(k_{x},\tfrac{\omega}{c})$ and $(\beta,\tfrac{\omega}{c})$ planes highlight unique features of hybrid ST modes that contrast with those of conventional modes. First, despite the finite spatial extent along $x$ for this pulsed field, the spectral projections are one-dimensional (1D), rather than 2D, because of the tight association between $k_{x}$ and $\omega$, and between $\beta$ and $\omega$. In contrast, the spatial and temporal degrees-of-freedom are usually separable for conventional pulsed beams [Fig.~\ref{Fig1}(a-d)]. Consequently, the field configurations associated with hybrid ST modes can be considered modal structures (each $\omega$ is assigned to a single axial wave number $\beta(\omega)$). Second, in contrast to the dispersion relationships for the planar-waveguide modes, those for hybrid ST modes are represented geometrically by a \textit{straight line}. The modal group index for $m\!=\!1$ is $\widetilde{n}_{1}\!=\!\cot{\theta_{1}}$, and the associated group velocity is $\widetilde{v}_{1}\!=\!c/\widetilde{n}_{1}\!=\!c\tan{\theta_{1}}$, which depends solely on the spectral tilt angle $\theta_{1}$ -- independently of the structure of the modal light-cone. Therefore, the group index is no longer restricted by the waveguide structure. Such versatility allows us to tune the group index for a hybrid ST mode by changing the field structure rather than modifying the waveguide. Moreover, the straight-line projection onto the $(\beta,\tfrac{\omega}{c})$-plane indicates that all orders of dispersion are eliminated across the whole spectrum. This remains the case even in presence of chromatic dispersion in the waveguide materials. Therefore, hybrid ST modes overcome both \textit{modal} and \textit{chromatic} dispersion inherent in the waveguide. 

\subsection{Coupling free-space ST wave packets to hybrid ST modes}

We synthesize ST wave packets in free space, which are then coupled into a planar waveguide to excite a hybrid ST mode. In free space ($n\!=\!1$), we produce a ST wave packet with a spectral tilt angle $\theta_{\mathrm{a}}$ and group velocity $\widetilde{v}_{\mathrm{a}}\!=\!c\tan{\theta_{\mathrm{a}}}$. When $\theta_{\mathrm{a}}\!>\!45^{\circ}$, the ST wave packet is superluminal $\widetilde{v}_{\mathrm{a}}\!>\!c$, and when $\theta_{\mathrm{a}}\!<\!45^{\circ}$, it is subluminal $\widetilde{v}_{\mathrm{a}}\!<\!c$. In both cases, the ST wave packet has a transverse spatial profile of width $\Delta x\!\approx\!\tfrac{\pi}{\Delta k_{x}}$, where $\Delta k_{x}$ is the spatial bandwidth. The spatio-temporal structure is such that each frequency $\omega$ is associated with a single spatial frequency according to \cite{Kondakci17NP,Kondakci19NC}:
\begin{equation}\label{Eq:parabola}
\frac{\Omega}{\omega_{\mathrm{o}}}=\frac{k_{x}^{2}}{2k_{\mathrm{o}}^{2}(1-\cot{\theta_{\mathrm{a}}})}.
\end{equation}
Because of this tight association between the spatial and temporal degrees of freedom, the spatial and temporal bandwidths, $\Delta k_{x}$ and $\Delta\lambda$, respectively, are related to each other. Indeed, for small bandwidths $\Delta\lambda\!\ll\!\lambda_{\mathrm{o}}$ we have
\begin{equation}\label{eq:bandwidth}
\frac{\Delta\lambda}{\lambda_{\mathrm{o}}}=\frac{(\Delta k_{x})^{2}}{2k_{\mathrm{o}}^{2}(1-\cot{\theta_{\mathrm{a}}})}.
\end{equation}
This formula indicates that the bandwidths $\Delta\lambda$ and $\Delta k_{x}$ are tightly related $\Delta\lambda\!\propto\!(\Delta k_{x})^{2}$, and thus also are the pulse width $\Delta t$ and transverse spatial width $\Delta x$. We show below, however, that this proportionality can be relaxed by exploiting a spectral recycling approach \cite{Hall21PRAspectralRecycling}.

Once coupled into the waveguide, the hybrid ST mode is identified by two numbers: (1) the modal index $m$ (whether TE or TM) of the associated conventional mode of the planar waveguide with transverse wave number $k_{y,m}$ along the confinement dimension $y$; and (2) its spectral tilt angle $\theta_{\mathrm{WG},m}$ inside the waveguide. Upon coupling to the $m^{\mathrm{th}}$-order waveguide mode, refraction at the waveguide entrance changes the spectral tilt angle in free space $\theta_{\mathrm{a}}$ to $\theta_{\mathrm{WG},m}$, such that the modal group index is $\widetilde{n}_{\mathrm{WG},m}\!=\!\cot{\theta_{\mathrm{WG},m}}$.

The distinction between superluminal and subluminal group velocities for free-space ST wave packets remains after coupling to hybrid ST modes. The free-space ST wave packet with $\theta_{\mathrm{a}}\!=\!45^{\circ}$ (a luminal plane-wave pulse \cite{Yessenov19PRA} with $k_{x}\!=\!0$) couples to the planar-waveguide mode having no spatial features along $x$, thereby corresponding to the light-lines in Fig.~\ref{Fig1}(b,d). The group index for these conventional modes is $\widetilde{n}_{m}$, which corresponds to the luminal limit in the waveguide. When $\theta_{\mathrm{a}}\!>\!45^{\circ}$, the superluminal ST wave packet couples to a hybrid ST mode with spectral tilt angle $\theta_{\mathrm{WG},m}$ such that $\widetilde{n}_{\mathrm{WG},m}\!=\!\cot{\theta_{\mathrm{WG},m}}\!<\!\widetilde{n}_{m}$; that is, the hybrid ST mode is also superluminal $\widetilde{v}_{\mathrm{WG},m}\!>\!\widetilde{v}_{m}$. Similarly, a subluminal ST wave packet couples to a subluminal hybrid ST mode with $\widetilde{v}_{\mathrm{WG},m}\!<\!\widetilde{v}_{m}$.

The change from the $\theta_{\mathrm{a}}$ for the free-space ST wave packet to $\theta_{\mathrm{WG},m}$ for the hybrid ST mode is governed by the law of refraction:
\begin{equation}\label{Eq:LawOfRefraction}
1-\widetilde{n}_{\mathrm{a}}=n\left\{n-(1-\eta_{m})\widetilde{n}_{\mathrm{WG},m}\right\},
\end{equation}
where $\widetilde{n}_{\mathrm{a}}\!=\!\cot{\theta_{\mathrm{a}}}$, $n$ is the refractive index of the waveguide material, and $\eta_{m}\!=\!\tfrac{1}{2}(\tfrac{k_{y,m}}{nk_{\mathrm{o}}})^{2}$ is a parameter resulting from the transverse wave number along $y$ associated with the planar-waveguide modes. This relationship, which is a consequence of the invariance of $k_{x}$ (the transverse wave number) and $\omega$ across a planar interface at normal incidence, generalizes to multiple modes the previously derived single-mode relationship \cite{Shiri20NC}. In our measurements here, we have focused on the two lowest-order modes of the waveguide. We switch between the two hybrid ST modes by tuning the coupling condition at the waveguide input. Therefore, the same ST wave packet in free space with spectral tilt angle $\theta_{\mathrm{a}}$ can couple in principle to any of the available hybrid ST modes, with modal group index determined by Eq.~\ref{Eq:LawOfRefraction}.

\section{Experiment}

\subsection{Synthesizing ST wave packets in free space}

\begin{figure}[t!]
\centering
\includegraphics[width=8.75cm]{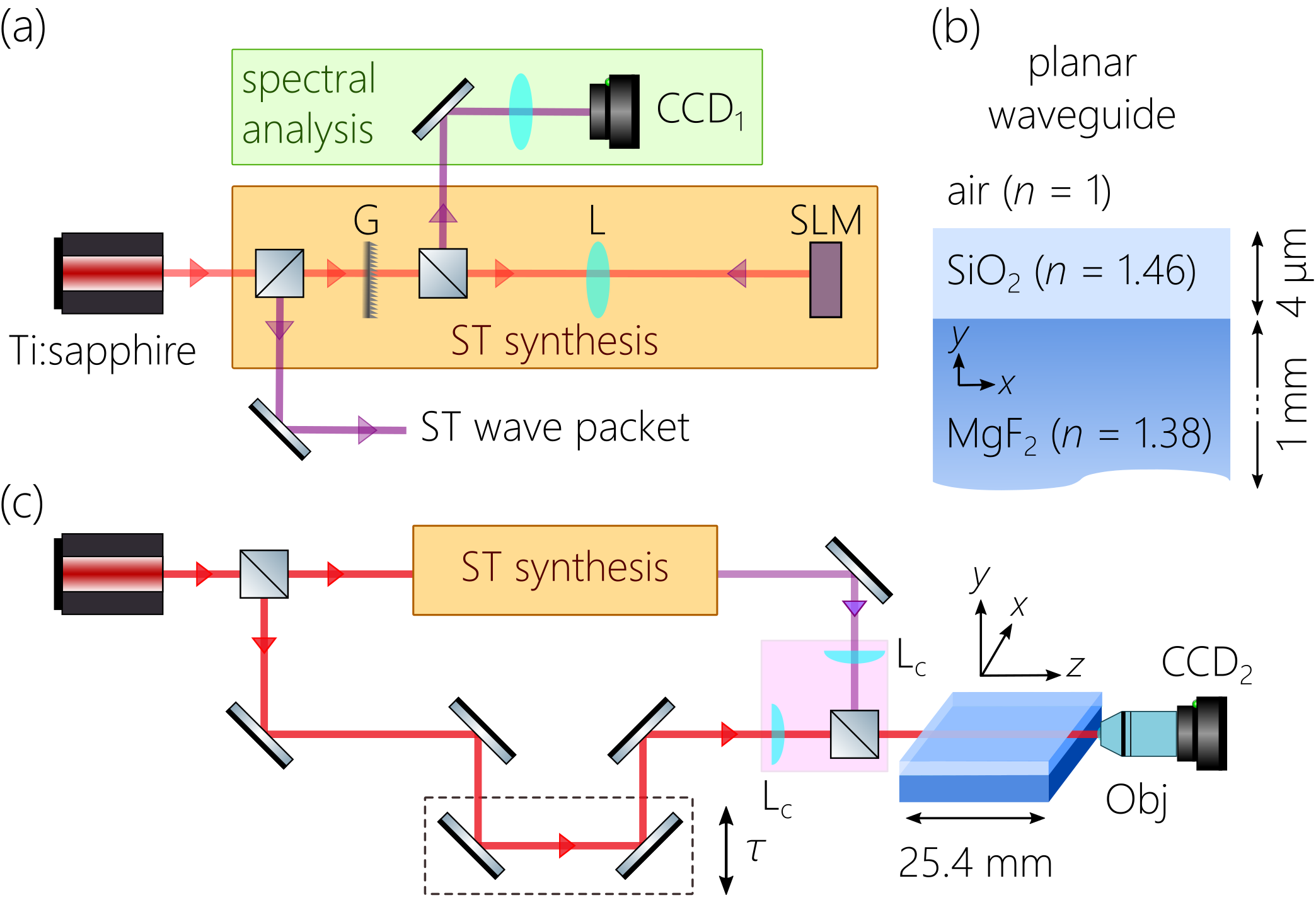}
\caption{\textbf{Experimental setup.} (a) ST wave packets are synthesized by first spectrally resolving pulses from a Ti:sapphire laser via a diffraction grating G and a cylindrical lens L. A SLM placed at the focal plane of L modulates the phase of the spectrally resolved wave front. The retro-reflected wave front returns to G to form the ST wave packet. The spatio-temporal spectrum is acquired by a CCD camera after a spatial Fourier transform. (b) Schematic of the planar waveguide cross section. (c) The ST synthesis setup from (a) is embedded in a Mach-Zehnder interferometer that delivers the ST wave packet along with a delayed reference pulse via a cylindrical lens L$_{\mathrm{c}}$ to the waveguide, whose output is imaged via an objective lens (Obj) to a CCD camera.}
\label{Fig2}
\end{figure}

We synthesize ST wave packets in free space using the pulsed-beam shaper depicted in Fig.~\ref{Fig2}(a), which takes the form of a folded $4f$ spectral-modulation scheme that is capable of inculcating arbitrary angular dispersion into a generic plane-wave pulse. Vertically polarized (along $y$) femtosecond pulses from a mode-locked Ti:sapphire laser (MIRA, Coherent; $\approx\!800$-nm central wavelength, $\approx\!6$-nm bandwidth, and $\approx\!150$-fs pulsewidth) are directed to a diffraction grating (1200~lines/mm) that spectrally resolves the pulse, and a cylindrical lens collimates the spectrum. A spatial light modulator (SLM; Hamamatsu X10468-02) modulates the phase of the collimated spectrally resolved wave front across a 0.7-nm bandwidth ($\sim\!3$-ps pulswidth). Each wavelength is deflected by a prescribed angle with respect to the $z$-axis to produce the target angular-dispersion profile $\varphi(\omega)$, and thus realize the desired relationship $k_{x}(\omega)\!=\!k\sin{\{\varphi(\omega)\}}$. The retro-reflected field from the SLM returns to the grating that reconstitutes the pulse and forms the ST wave packet.

Here, the angular dispersion is necessary along one transverse dimension only (the unbounded $x$-dimension of the waveguide), while leaving the field unchanged along the $y$-dimension (where light will be focused along the bounded dimension of the planar waveguide). The structure of the synthesized spatio-temporal spectrum is verified by implementing a spatial Fourier transform along $x$ for the spectrally resolved wave front retro-reflected from the SLM [Fig.~\ref{Fig2}(a)]. This measurement reveals the spectral projection onto the $(k_{x},\tfrac{\omega}{c})$-plane, which we refer to as $k_{x}(\lambda)$. From this measurement, we can estimate the axial wave number $\beta(\lambda)$ in the waveguide.

\subsection{Coupling free-space ST wave packets to the waveguide}

The fabricated waveguide consists of a 4-$\mu$m-thick layer of silica ($n\!\approx\!1.46$ at $\lambda\!\sim\!800$~nm) deposited by e-beam evaporation on a 1-mm-thick, $25.5\times25.5$-mm$^2$ area MgF$_2$ substrate ($n\!\approx\!1.38$ at $\lambda\!\sim\!800$~nm); see Fig.~\ref{Fig2}(b). The guiding silica layer is air-clad from the top. This waveguide has five guided modes, of which we excite the first two index-guided TM modes (TM$_0$ and TM$_1$, so that we have $m\!=\!0,1$). The synthesised ST wave packet is focused by a cylindrical lens L$_{\mathrm{c}}$ along $y$ (the bounded dimension of the planar waveguide) to the input waveguide facet that is placed at its focal plane. Switching between the first two modes is achieved by displacing the waveguide vertically along $y$. The output field profile is imaged onto a CCD camera with an objective lens (Olympus PLN $20\times$); see Fig.\ref{Fig2}(c). 

\subsection{Measuring the modal group delay}

In our previous work on hybrid ST wave packets \cite{Shiri20NC}, we estimated the group velocity of the free-space ST wave packet from the measured spatio-temporal spectrum and then deduced the group index in the waveguide from the law of refraction in Eq.~\ref{Eq:LawOfRefraction}. This is indirect estimation of the group index of the hybrid ST modes. Here, we implement a new experimental strategy to measure the modal group delay in the waveguide \textit{directly} by interfering the hybrid ST mode with a simultaneously coupled generic pulsed field. This is crucial to obtain accurate estimates of each hybrid ST mode separately. We place the free-space spatio-temporal synthesis setup [Fig.~\ref{Fig2}(a)] in one arm of a Mach-Zehnder interferometer, and place an optical delay $\tau$ in the other arm that is traversed by the $\approx\!150$-fs reference pulses from the Ti:sapphire laser [Fig.~\ref{Fig2}(c)].

We first synchronize the free-space ST wave packet (3-ps pulsewidth, $\widetilde{v}_{\mathrm{a}}\!=\!c\tan{\theta_{\mathrm{a}}}$) with the reference pulse (150-fs pulsewidth, $\widetilde{v}_{\mathrm{ref}}\!=\!c$) at the focal plane of the lens L$_{\mathrm{c}}$ that couples the field to the waveguide. Overlap in space and time of these two wave packets is revealed through the observation of high-visibility spatially resolved interference fringes. This condition provides the baseline for the modal group-delay measurement. The group velocity of the reference pulse in the waveguide is $\widetilde{v}_{\mathrm{ref}}\!=\!c/\widetilde{n}_{m}$, where $\widetilde{n}_{m}$ is the modal group index of the $m^{\mathrm{th}}$-order conventional planar-waveguide mode (which can be calculated from the waveguide structure), and that of the associated hybrid ST mode is $\widetilde{v}_{\mathrm{WG},m}\!=\!c\tan{\theta_{\mathrm{WG},m}}$. At the output, the interference fringes are lost because of the relative delay introduced between the two wave packets after traversing the waveguide $\Delta\tau_{m}\!=\!\tfrac{L}{c}(\widetilde{n}_{\mathrm{WG},m}-\widetilde{n}_{m})$, where $L$ is the waveguide length. Of course, $\Delta\tau_{m}$ is positive for subluminal wave packets (the hybrid ST mode emerges from the waveguide after the conventional mode), and $\Delta\tau_{m}$ is negative for superluminal wave packets (the hybrid ST mode emerges first). We adjust the optical delay $\tau$ in the reference arm to compensate for this relative group delay between these two pulses, thus revealing high-visibility interference at the output. From $\tau$ we estimate the group delay accrued by the hybrid guided ST mode $\tfrac{L}{c}\widetilde{n}_{\mathrm{WG},m}$, and thence obtain its modal group index $\widetilde{n}_{\mathrm{WG},m}$. 

\section{Tuning the group index of hybrid ST modes}

\begin{figure*}[ht!]
\centering
\includegraphics[width=15.5cm]{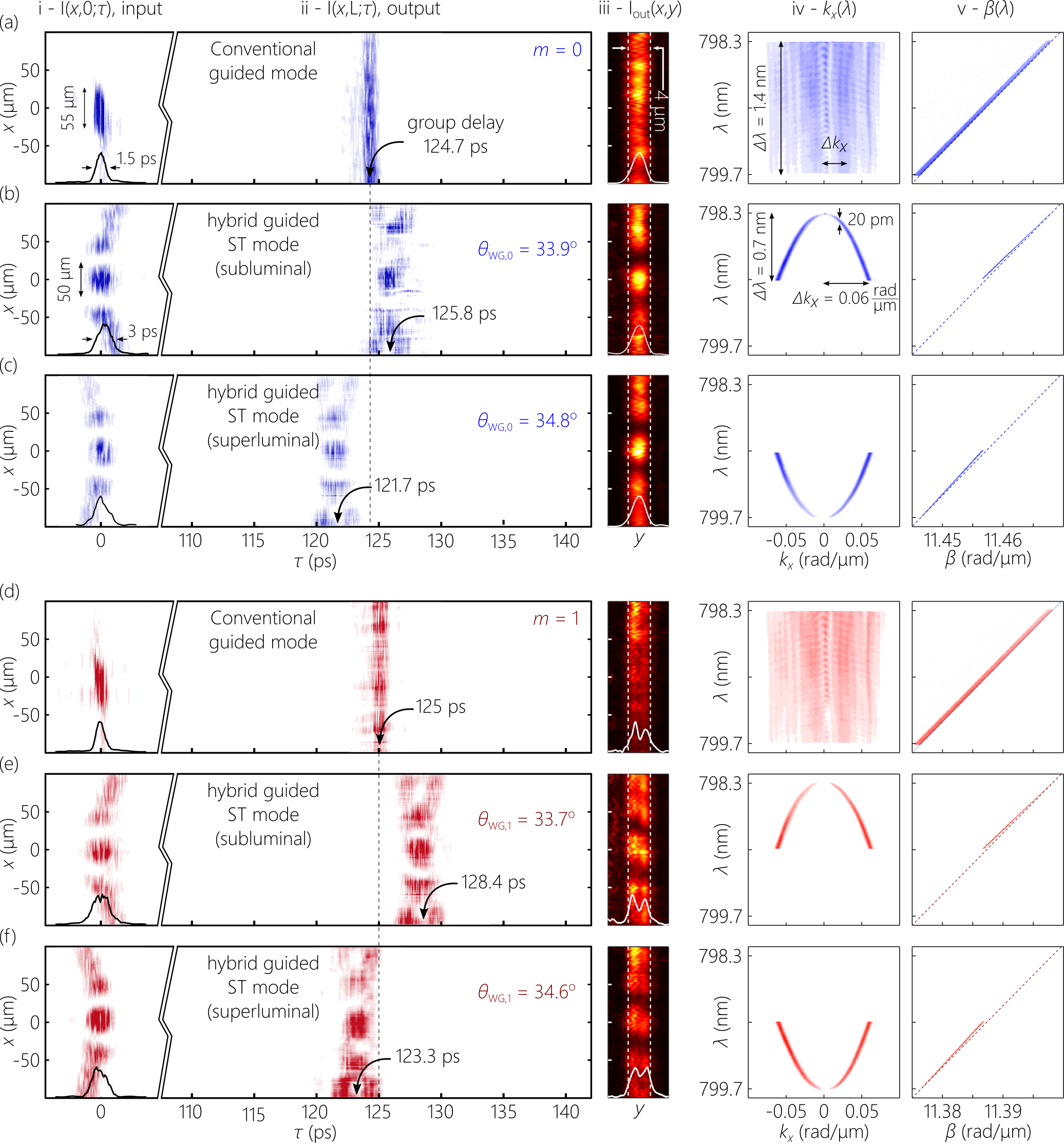}
\caption{\textbf{Measurements of the group delay for conventional modes and hybrid ST modes in a multimode planar waveguide.} For each mode, we plot the following measurements (from left to right): (i) the input spatio-temporal intensity profile $I(x,z=0;t)$; (ii) the output profile $I(x,z=L;t)$; (iii) the time-averaged output intensity $I_{\mathrm{out}}(x,y)$, with $I_{\mathrm{out}}(0,y)$ plotted at the bottom; (iv) the angular dispersion $k_{x}(\lambda)$; and (v) the dispersion relationship $\beta(\lambda)$. The two vertical dashed white lines in (iii) delimit the waveguide interfaces, and the dotted line in (v) is the waveguide light-line.(a) The fundamental mode $m\!=\!0$. The input field is focused along $x$ and diffracts with propagation. (b) A subluminal hybrid ST mode with $m\!=\!0$ and $\theta_{\mathrm{a}}\!=\!44^{\circ}$ is invariant along the waveguide. (c) Same as (b) except that the hybrid ST mode is superluminal with $\theta_{\mathrm{a}}\!=\!46^{\circ}$. (d-f) Same as (a-c) for $m\!=\!1$.}
\label{Fig3}
\end{figure*}

We first determine the baseline for the modal group-delay measurements. We couple a 1.5-ps pulse (1.4-nm-bandwidth, obtained by idling the SLM, $\theta_{\mathrm{a}}\!=\!45^{\circ}$ and $\widetilde{n}_{\mathrm{a}}\!=\!1$) after focusing it to a spot size of width $\approx\!55$~$\mu$m along the unconstrained dimension $x$ to excite the first-order mode $m\!=\!0$. The input wave packet has finite spatial \textit{and} temporal bandwidths corresponding to the scenario sketched in Fig.~\ref{Fig1}(a,b). We reconstruct the spatio-temporal intensity profile at the waveguide input after synchronizing the 1.5-ps wave packets at the waveguide entrance with the 150-fs plane-wave reference pulses from the laser \cite{Kondakci19NC,Yessenov19OE} [Fig.~\ref{Fig3}(a)-i]. Because both pulses travel at the \textit{same} group velocity along the waveguide, they emerge at the output while \textit{still} synchronized [Fig.~\ref{Fig3}(a)-ii]. The modal group index for $m\!=\!0$ is $\widetilde{n}_{0}\!\approx\!1.463$ ($\theta_{0}\!=\!34.3^{\circ}$), resulting in a group delay of $\approx\!124.7$~ps after the 25.5-mm-long waveguide. However, the initially focused input field structure along $x$ is lost when it reaches the waveguide output, and the time-averaged intensity profile $I_{\mathrm{out}}(x,y)$ at the output exhibits clear diffractive spreading along $x$ [Fig.~\ref{Fig3}(a)-iii]. The measured spatio-temporal spectrum $k_{x}(\lambda)$ is a 2D domain that is separable along $k_{x}$ and $\lambda$, as expected, and the spectral projection of the dispersion relationship $\beta(\lambda)$ is also a 2D domain that lies in the immediate vicinity of the light-line.

We next consider a \textit{subluminal} ST wave packet synthesized in free space with $\theta_{\mathrm{a}}\!=\!44^{\circ}$ ($\widetilde{n}_{\mathrm{a}}\!=\!1.036$) and $\approx\!0.7$-nm-bandwidth (pulsewidth 3~ps), which is launched into the waveguide to excite the same fundamental mode $m\!=\!0$ along $y$ as above. The spatio-temporal profile of the ST wave packet at a fixed axial plane $z$ and fixed coordinate $y$ has a characteristic X-shape in terms of $x$ and $t$ [Fig.~\ref{Fig3}(b)-i] \cite{Kondakci19NC,Yessenov19OE}. The hybrid ST mode retains its initial profile without diffractive or dispersive spreading after traversing the waveguide, while incurring a relative group delay of $\approx\!1.1$~ps with resepct to the conventional $m\!=\!0$ mode [Fig.~\ref{Fig3}(b)-ii]. Here, $k_{x}(\lambda)$ [Fig.~\ref{Fig3}(b)-iv] shows a profound departure from that of the conventional mode: rather than a 2D domain in $(k_{x},\lambda)$-space, we now have a 1D trajectory that can be approximated by a parabola in the vicinity of $k_{x}\!=\!0$ (Eq.~\ref{Eq:parabola}). In other words, each spatial frequency $\pm k_{x}$ is associated with a single wavelength $\lambda$, and the dispersion relationship $\beta(\lambda)$ is now a straight line, indicating the absence of dispersion. The spectral projection onto the $(\beta,\tfrac{\omega}{c})$-plane makes an angle $\theta_{\mathrm{WG},m}\!=\!33.9^{\circ}$ with the $\beta$-axis, indicating a group modal index of $\widetilde{n}_{\mathrm{WG},0}\!\approx\!1.487$, which agrees with that estimated from the measured group delay.

\begin{figure}[t!]
\centering
\includegraphics[width=8cm]{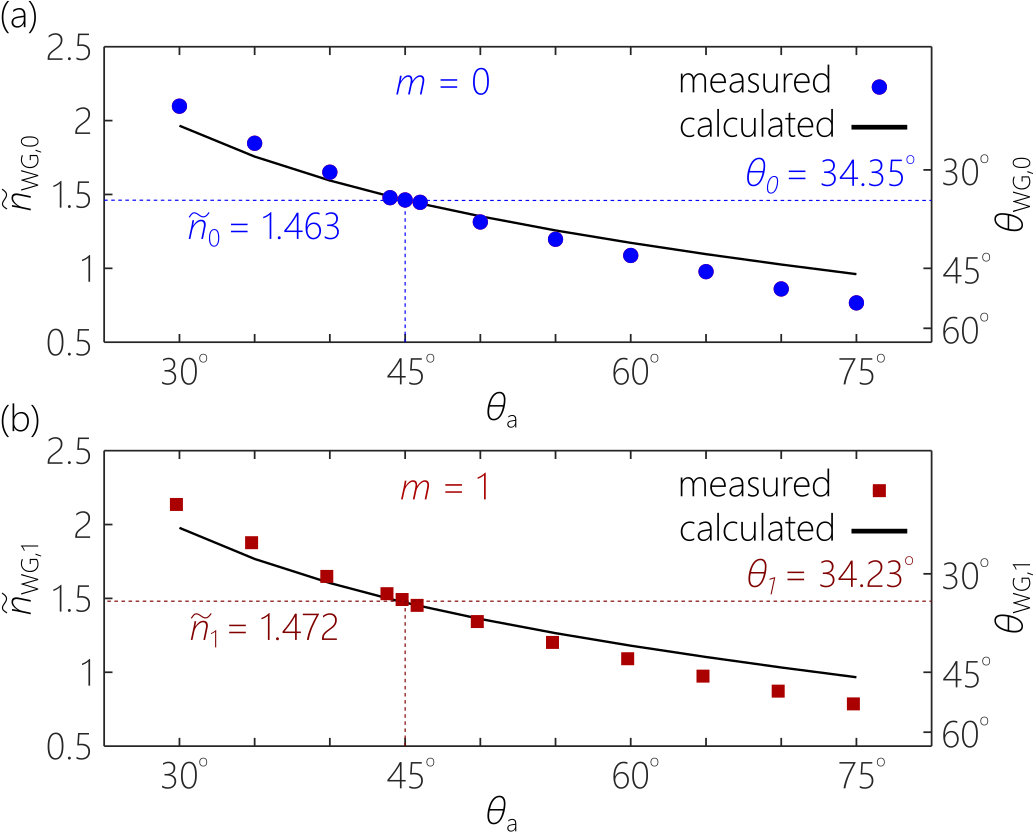}
\caption{\textbf{Tuning the group index for hybrid ST modes.} (a) Measured group index $\widetilde{n}_{\mathrm{WG},0}$ of the hybrid ST mode corresponding to the fundamental planar-waveguide mode over the span $30^{\circ}\!<\!\theta_{\mathrm{a}}\!<\!70^{\circ}$. The points are data and the solid curve is the theoretical prediction based on Eq.~\ref{Eq:LawOfRefraction}. (b) Same as (a) for $\widetilde{n}_{\mathrm{WG},1}$ corresponding to the second-order planar-waveguide mode.}
\label{Fig4}
\end{figure}

Unlike conventional modes whose group indices at any wavelength are determined by the waveguide itself, the group index of a hybrid ST mode can be tuned continuously by varying the spatio-temporal field structure, thereby changing $\theta$. This is further confirmed in Fig.~\ref{Fig3}(c) where we plot the measurement results for a \textit{superluminal} hybrid ST mode with $\theta_{\mathrm{a}}\!=\!46^{\circ}$. We retain the same spatial and temporal bandwidths used in the subluminal mode described above, so that the modal width and pulsewidth are unchanged. However, the relative group delay incurred with resepct to the conventional mode [Fig.~\ref{Fig3}(a)] upon traversing the waveguide is $\approx\!-3$~ps. The sign of the curvature of $k_{x}(\lambda)$ [Fig.~\ref{Fig3}(c)-iv] has switched with respect to its subluminal counterpart, and the spectral projection onto the $(\beta,\tfrac{\omega}{c})$-plane [Fig.~\ref{Fig3}(c)-v] makes an angle $34.8^{\circ}$ with the $\beta$-axis, $\widetilde{n}_{\mathrm{WG},0}\!\approx\!1.439$.

We plot in Fig.~\ref{Fig4}(a) the measured group index $\widetilde{n}_{\mathrm{WG},0}$ for the hybrid ST mode with $m\!=\!0$ as we vary $\theta_{\mathrm{a}}$ for the free-space ST wave packet. When $\theta_{\mathrm{a}}\!=\!45^{\circ}$, the ST wave packet is a plane-wave pulse, and $\widetilde{n}_{\mathrm{WG},0}\!=\!\widetilde{n}_{0}\!=\!1.463$, corresponds to the conventional mode; when $\theta_{\mathrm{a}}\!<\!45^{\circ}$, the wave packet is subluminal $\widetilde{n}_{\mathrm{WG},0}\!>\!\widetilde{n}_{0}$; and when $\theta_{\mathrm{a}}\!>\!45^{\circ}$, the wave packet is superluminal $\widetilde{n}_{\mathrm{WG},0}\!<\!\widetilde{n}_{0}$. We have spanned the range $30^{\circ}\!<\!\theta_{\mathrm{a}}\!<\!75^{\circ}$, corresponding to $0.96\!<\!\widetilde{n}_{\mathrm{WG},0}\!<\!1.97$, which is an unprecedented tunability range achieved without changing the waveguide structure itself -- only tailoring the spatio-temporal structure of the input field. Consequently, $\widetilde{n}_{\mathrm{WG},0}$ is now untethered to the baseline value $\widetilde{n}_{0}$ of the underlying conventional mode.

We repeated these measurements for the second-order mode ($m\!=\!1$) whose modal group index is $\widetilde{n}_{1}\!\approx\!1.472$ [Fig.~\ref{Fig3}(d)]. We can tune the modal group index $\widetilde{n}_{\mathrm{WG},1}$ for the hybrid ST mode with $m\!=\!1$ to subluminal [Fig.~\ref{Fig3}(e)] or superluminal [Fig.~\ref{Fig3}(f)] values, similarly to what we did for the first-order mode. Indeed, we can even render this higher-order mode significantly faster than the fundamental mode. We plot in Fig.~\ref{Fig4}(b) measurements of $\widetilde{n}_{\mathrm{WG},1}$, showing the same tunability range as for the fundamental mode while varying $\theta_{\mathrm{a}}$. Once again, $\widetilde{n}_{\mathrm{WG},1}$ is untethered to the baseline value $\widetilde{n}_{1}$ of the planar-waveguide conventional mode $m\!=\!1$. Because this same effect can be implemented with any other mode of the planar waveguide, we have therefore severed the link between the modal group index $\widetilde{n}_{\mathrm{WG},m}$ and the modal order $m$.

\section{Tuning the size of the hybrid ST mode}

Because of the tight association between the spatial and temporal degrees of freedom of a ST wave packet as determined by the spectral tilt angle $\theta_{\mathrm{a}}$, changing $\theta_{\mathrm{a}}$ at a fixed temporal bandwidth $\Delta\lambda$ results unavoidably in a concomitant change in the spatial bandwidth $\Delta k_{x}$, and thus the modal width $\Delta x$ (Eq.~\ref{eq:bandwidth}). The modal profile along $y$, on the other hand, is fixed and determined by the waveguide structure. Changing $\theta_{\mathrm{a}}$ to equalize the modal group indices at the same $\Delta\lambda$ changes the modal widths along $x$. We demonstrate the versatility of hybrid ST modes by verifying that their width $\Delta x$ can be varied independently of $\theta_{\mathrm{a}}$.

\begin{figure*}[t!]
\centering
\includegraphics[width=15cm]{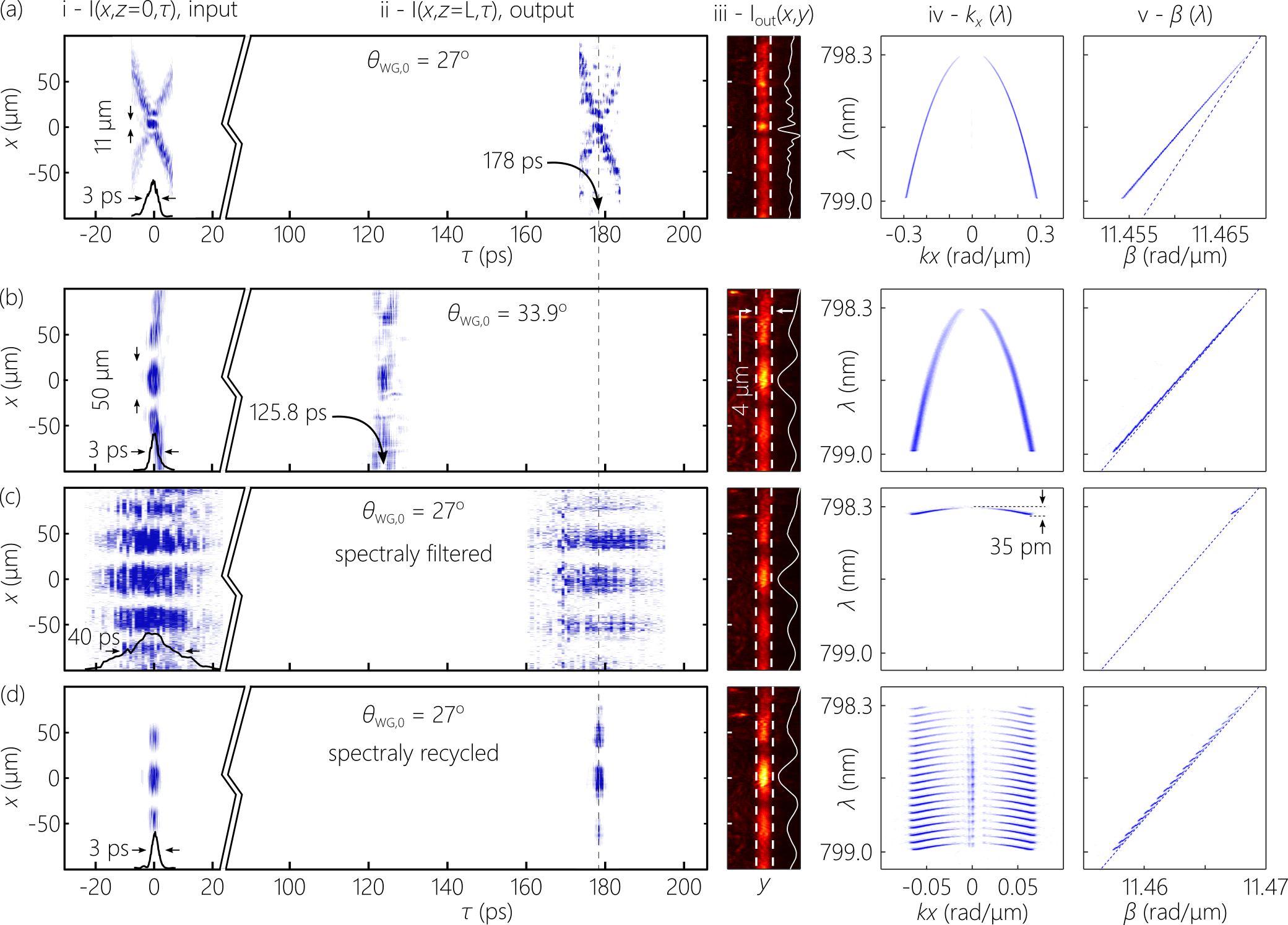}
\caption{\textbf{Varying the spatial width of hybrid ST modes -- at fixed bandwidth and spectral tilt angle -- via spectral recycling.} The panels in each row are arranged similarly to Fig.~\ref{Fig3}, and we add $I_{\mathrm{out}}(x,y=0)$ in (iii). (a) Subluminal hybrid ST mode with $m\!=\!0$, $\theta_{\mathrm{a}}\!=\!30^{\circ}$, $\theta_{\mathrm{WG},0}\!=\!27^{\circ}$, $\Delta\lambda\!=\!0.7$~nm, modal width $\Delta x\!\approx\!11$~$\mu$m, and pulsewidth $\Delta\tau\!\approx\!3$~ps. (b) Subluminal hybrid ST mode with $m\!=\!0$, $\theta_{\mathrm{a}}\!=\!44^{\circ}$, $\theta_{\mathrm{WG},0}\!=\!33.9^{\circ}$, $\Delta\lambda\!=\!0.7$~nm, $\Delta x\!\approx\!50$~$\mu$m, and $\Delta\tau\!\approx\!3$~ps. (c) Same as (a), but spectrally filtered to $\Delta\lambda\!=\!35$~pm, $\Delta\tau\approx\!40$~ps, and $\Delta x\!\approx\!50$~$\mu$m. (d) Same as (c), but spectrally recycled so that we restore $\Delta\lambda\!=\!0.7$~nm and $\Delta\tau\!\approx\!3$~ps, while maintaining $\Delta x\!\approx\!50$~$\mu$m.}
\label{Fig5}
\end{figure*}

Specifically, we consider here a strategy for maintaining $\Delta x$ fixed as we change $\theta_{\mathrm{a}}$ at fixed $\Delta\lambda$. We plot in Fig.~\ref{Fig5}(a) measurements for a subluminal hybrid ST mode at $\theta_{\mathrm{a}}\!=\!30^{\circ}$, $\theta_{\mathrm{WG},0}\!=\!27^{\circ}$, and $\widetilde{n}_{\mathrm{WG},0}\!=\!1.963$. Maintaining $\Delta\lambda\!=\!0.7$~nm as in Fig.~\ref{Fig3}, the modal width becomes only $\Delta x\!\approx\!11$~$\mu$m [Fig.~\ref{Fig5}(a)], in comparison to $\Delta x\!\approx\!50$~$\mu$m at $\theta_{\mathrm{a}}\!=\!44^{\circ}$, $\theta_{\mathrm{WG},0}\!=\!33.9^{\circ}$, and $\widetilde{n}_{\mathrm{WG},0}\!=\!1.487$ [Fig.~\ref{Fig5}(b)]. According to Eq.~\ref{eq:bandwidth}, realizing a modal width of $\Delta x\!\approx\!50$~$\mu$m at $\theta_{\mathrm{a}}\!=\!30^{\circ}$ requires a reduction in bandwidth to a mere $\Delta\lambda\!\approx\!35$~pm, and thus increasing the pulsewidth to 40-ps [Fig.~\ref{Fig5}(c)]. However, spectral recycling \cite{Hall21PRAspectralRecycling} overcomes this difficulty by allowing us to increase $\Delta\lambda$ in Fig.~\ref{Fig5}(c) \textit{without} changing $\Delta k_{x}$. This is achieved by recycling the same spatial frequency at multiple wavelengths [Fig.~\ref{Fig5}(d)-iv,v]. We divide the target bandwidth of $\Delta\lambda\!=\!0.7$~nm into $N\!=\!20$ segments of bandwidth $\Delta\lambda\!=\!35$~pm each. Because no new spatial frequencies are added, the width remains $\Delta x\!\approx\!50$~$\mu$m. However, increasing $\Delta\lambda$ to 0.7~nnm reduces the pulsewidth to 3~ps at $\theta_{\mathrm{a}}\!=\!30^{\circ}$. The projected dispersion relationship takes the form of 20 displaced but parallel segments, each having the same spectral tilt angle $\theta_{\mathrm{WG},0}\!=\!27^{\circ}$ with the $\beta$-axis. Using this spatial recycling strategy, we can tailor the modal width along $x$ to match that of any other mode at a different spectral tilt angle but same temporal bandwidth $\Delta\lambda$.

\section{Discussion}

It is important to point out here the unique nature of angular dispersion undergirding hybrid ST modes, which allows for their unique features that distinguish them from conventional modes. In a non-dispersive medium, the dispersion relationship is $k\!=\!n\tfrac{\omega}{c}$, whereupon optical pulses do not experience temporal broadening. Optical confinement in a waveguide inevitably enforces a departure from this linear relationship, and thus the appearance of GVD and higher-order dispersion terms. The new linear dispersion relationship $\Omega\!=\!(\beta-nk_{\mathrm{o}})\widetilde{v}$ imposed by the spatio-temporal spectral structure of a ST wave packet circumvents this seemingly inescapable conclusion. However, realization of this new linear dispersion relationship requires \textit{non-differentiable} angular dispersion \cite{Hall21OL,Hall21OL3NormalGVD,Hall21OE1NonDiff}; that is, the derivative of the propagation angle $\varphi(\omega)$ is \textit{not} defined at $\omega\!=\!\omega_{\mathrm{o}}$. Indeed, in the small-angle approximation it can be shown that $\varphi(\omega)\!\approx\!\eta\sqrt{\tfrac{\Omega}{\omega_{\mathrm{o}}}}$, which is \textit{not} differentiable at $\omega\!=\!\omega_{\mathrm{o}}$ \cite{Hall21OL}; here $\widetilde{n}\!=\!1-\tfrac{1}{2}\eta^{2}$, and $\eta$ is a frequency-independent constant. This can also be seen clearly in Fig.~\ref{Fig1}(f,h) when the spectral trajectory at the intersection of the tilted spectral plane with the modal light-cone reaches the point at $\omega\!=\!\omega_{\mathrm{o}}$. The projected spectral trajectory plotted with $\omega$ as the horizontal axis has a slope that becomes infinite as $\omega\!\rightarrow\!\omega_{\mathrm{o}}$. This requisite non-differentiable angular dispersion cannot be produced via conventional optical components, such as diffraction gratings or prisms \cite{Hall21OL3NormalGVD,Hall21OE1NonDiff}. However, the spatio-temporal spectral modulation strategy depicted in Fig.~\ref{Fig2}(a) serves as a universal angular-dispersion synthesizer \cite{Hall21OE2Universal}, and is capable of producing the requisite non-differentiable angular dispersion along $x$. This form of non-differentiable angular dispersion enables tunable modal group index and ensures the absence of GVD for hybrid ST modes. We expect further developments regarding this nascent concept of non-differentiable angular dispersion. 

There are several new avenues to be investigated with hybrid ST modes that make use of recent developments in their freely propagating counterparts; specifically with regards to the potential for controlling the axial evolution of the field. First, ST wave packets that undergo tunable GVD in free space have been demonstrated \cite{Yessenov21ACSP}. Such wave packets will produce dispersion-free hybrid ST modes in planar waveguide comprising a dispersive material. Second, axial spectral encoding of ST wave packets allows for the on-axis spectrum to controllably evolve along the axis \cite{Motz21PRA}. Endowing hybrid ST modes with this feature may be useful in enhancing nonlinear parametric interactions. Third, by replacing the one-dimensional spectral trajectory on the modal light-cone surface with a precisely sculpted two-dimensional domain, one can produce \textit{accelerating} hybrid ST modes \cite{Yessenov20PRL2,Hall21OL4Acceleration}; i.e., the group index of the mode changes along the propagation axis $\widetilde{n}_{\mathrm{WG},m}\!=\!\widetilde{n}_{\mathrm{WG},m}(z)$.

In hybrid ST modes, the field interacts with the planar waveguide structure along one dimensions $y$ and the spatio-temporal structure is sculpted along the unbounded or free dimension $x$. The unique features of these novel modes are a consequence of the angular dispersion along $x$ overriding the constraints imposed by the boundary conditions along $y$. It is important to distinguish between such hybrid ST modes and another class of field configurations that can be referred to as `guided ST supermodes' in which ST wave packets are coupled into conventional waveguides or fibers that confine light in both dimensions. In this scenario, the ST wave packet interacts with the waveguide boundary conditions in both dimensions, and the wave packet spectrum is discretized and restricted to a superposition of the waveguide modes but with the wave number of each mode assigned to a particular wavelength. These fields have been studied theoretically \cite{Zamboni01PRE,Zamboni02PRE,Zamboni03PRE} by exploiting a particular class of ST wave packets known as X-waves \cite{Lu92IEEEa,Saari97PRL}. To the best of our knowledge, guided ST supermodes have \textit{not} yet been realized experimentally. Recent theoretical studies have re-examined guided ST supermodes by investigating the potential for synthesizing them within the waveguide itself, either by exploiting time-varying refractive indices to induce indirect photonic transitions from a single waveguide mode to a target guided ST supermode \cite{Guo21PRR}, or through high-energy nonlinear interactions \cite{Kibler21pRL}, which may lead to the realization of so-called spatio-temporal `helicon' wave packets \cite{Bejot21ACSP} after incorporating orbital angular momentum into the field. Another related effort has been to study the propagation of focus-wave modes \cite{Brittingham83JAP} in optical fibers \cite{Vengsarkar92JOSAA}, and more recently to investigate the impact of the presence of orbital angular momentum on such guided fields \cite{Ruano21JO}.

\section{Conclusion}

In conclusion, we have demonstrated that the recently developed concept of hybrid guided ST modes in single-mode planar waveguides can be extended to the multimode regime. By sculpting the spatio-temporal spectral structure of the field along the unconstrained waveguide dimension, we construct field configurations that are confined in both dimensions, and whose characteristics deviate dramatically from those of the conventional modes upon which they are based. Specifically, tuning the spatio-temporal field structure enables us to override the impact of the boundary conditions of the waveguide and vary the group index of the dispersion-free hybrid ST mode independently of the waveguide structure. In the context of a multimode planar waveguide, higher-order modes usually have larger group indices than their lower-order counterparts, and thus accrue larger group delays. In contrast, because the group index of a hybrid ST mode is no longer tethered to the group index of the waveguide mode, the group indices of different modes can be tuned independently, thereby leading to a reversal of the conventional scenario: higher-order modes can now travel with the same or lower group indices than lower-order modes. Furthermore, we exploit the concept of spectral recycling to circumvent the necessary proportionality between the transverse modal width of the hybrid ST mode and its pulse linewidth. This feature provides control over the modal width at constant modal group index \textit{and} pulsewidth. These results are expected to be useful in engineering multi-wavelength nonlinear optical interactions in a multimode waveguide.

\section*{Acknowledgments}
The authors thank M. Oberdzinsky and J. Cha (Optonetics, Inc.) for deposition of the silica layer used here.This work was supported by the U.S. Office of Naval Research (ONR) under contracts N00014-17-1-2458 and N00014-20-1-2789.

\bibliography{diffraction}

\end{document}